\documentclass[bib]{statapress}

\usepackage[crop,newcenter,frame]{pagedims}

\usepackage{sj}

\usepackage{stata}

\usepackage{shadow}

\usepackage{bbm}

\usepackage{fancyvrb}

\sjsetissue{$vv$}{$ii$}{$mm$}{$yyyy$}

\usepackage{amssymb,amsmath,bbm,mathrsfs,longtable,graphicx,subcaption,enumitem}
\usepackage{hyperref}
\hypersetup{colorlinks=true,       
    linkcolor=blue,          
    citecolor=blue,        
    filecolor=magenta,      
    urlcolor=blue           
}
\newcommand{\E}{\mathbb{E}}

\newcommand{\I}{\mathbbm{1}}
\DeclareMathOperator*{\argmin}{arg\,min}

\newcommand{\bW}{\mathbf{W}}

\newcommand{\bZ}{\mathbf{Z}}

\newcommand{\be}{\mathbf{e}}

\newcommand{\bw}{\mathbf{w}}

\newcommand{\balpha}{\boldsymbol{\alpha}}

\newcommand{\bbeta}{\boldsymbol{\beta}}

\newcommand{\bgamma}{\boldsymbol{\gamma}}

\newcommand{\bxi}{\boldsymbol{\xi}}

\newcommand{\blambda}{\boldsymbol{\lambda}}

\begin{document}

\inserttype[st0001]{article}

\title[Heterogeneous RD Treatment Effects]{rdhte: Conditional Average Treatment Effects in RD Designs}

\author{S. Calonico, M. D. Cattaneo, M. H. Farrell, F. Palomba and R. Titiunik}{
  Sebastian Calonico\\UC Davis\\ Davis, CA\\scalonico@ucdavis.edu
  \and
  Matias D. Cattaneo\\Princeton University\\Princeton, NJ\\cattaneo@princeton.edu
  \and
  Max H. Farrell\\UC Santa Barbara\\Santa Barbara, CA\\mhfarrell@gmail.com
  \and
  Filippo Palomba\\Princeton University\\Princeton, NJ\\fpalomba@princeton.edu
  \and
  Rocio Titiunik\\Princeton University\\Princeton, NJ\\titiunik@princeton.edu
}

\maketitle

\begin{abstract}
	Understanding causal heterogeneous treatment effects based on pretreatment covariates is a crucial aspect of empirical work. Building on \cite{Calonico-Cattaneo-Farrell-Palomba-Titiunik_2025_wp}, this article discusses the software package \texttt{rdhte} for estimation and inference of heterogeneous treatment effects in sharp regression discontinuity (RD) designs. The package includes three main commands: \texttt{rdhte} conducts estimation and robust bias-corrected inference for heterogeneous RD treatment effects, for a given choice of the bandwidth parameter; \texttt{rdbwhte} implements automatic bandwidth selection methods; and \texttt{rdhte\_lincom} computes point estimates and robust bias-corrected confidence intervals for linear combinations, a post-estimation command specifically tailored to \texttt{rdhte}. We also provide an overview of heterogeneous effects for sharp RD designs, give basic details on the methodology, and illustrate using an empirical application. Finally, we discuss how the package \texttt{rdhte} complements, and in specific cases recovers, the canonical RD package \texttt{rdrobust} \citep{Calonico-Cattaneo-Farrell-Titiunik_2017_Stata}.\medskip
	
\end{abstract}

\newpage

\section{Introduction}

Studying causal heterogeneous treatment effects based on pretreatment covariates is an essential part of modern empirical work, as it helps uncover fairness concerns, differential impacts, and informs targeted policy interventions. While the regression discontinuity (RD) design has become a widely used tool for causal inference, existing methods primarily focus on estimating a single overall average treatment effect, leaving a gap in rigorous approaches for studying conditional average treatment effects. \cite{Calonico-Cattaneo-Farrell-Palomba-Titiunik_2025_wp} addressed this gap by developing a unified, theoretically grounded methodology for heterogeneity analysis in RD designs, allowing researchers to systematically examine how treatment effects vary across subpopulations based on pretreatment characteristics. \cite{Cattaneo-Idrobo-Titiunik2019_book,Cattaneo-Idrobo-Titiunik2023_book} provide a practical introduction to RD designs, estimation, and inference. For a review of recent literature on RD, see \cite{Cattaneo-Titiunik_2022_ARE}. 

This article discusses the general-purpose software package \texttt{rdhte}, which implements the main methodological results in \cite{Calonico-Cattaneo-Farrell-Palomba-Titiunik_2025_wp}. The package is composed by three commands:

\begin{itemize}
    \item \texttt{rdhte}. Given a choice of bandwidth for localization near the cutoff determining treatment assignment, this command implements local polynomial regression estimation and inference methods allowing for interactions with pretreatment covariates. More precisely, estimation is implemented using a local weighted least squares regression incorporating semi-linear interactions with the covariates used for heterogeneity analysis, while inference is conducted using robust bias-correction \citep{Calonico-Cattaneo-Titiunik_2014_ECMA,Calonico-Cattaneo-Farrell_2018_JASA,Calonico-Cattaneo-Farrell_2022_Bernoulli}. The package also allows for inclusion of other pretreatment covariates for efficiency gains purposes \citep{Calonico-Cattaneo-Farrell-Titiunik_2019_RESTAT}. A noteworthy feature of this command is that its implementation relies on base commands ({\tt regress} in \texttt{Stata}) for generic linear-in-parameters least squares estimation and inference, making \texttt{rdhte} more numerically efficient and robust. This feature implies that {\tt rdhte} may not perfectly match {\tt rdrobust} (see Section \ref{sec: Comparison with rdrobust}), and may not yield identical numerical results across different statistical software platforms.
    
    \item \texttt{rdbwhte}. This command implements bandwidth selection tailored towards inclusion of covariates for heterogeneity analysis and efficiency improvements. See \cite{Calonico-Cattaneo-Farrell_2020_ECTJ} for a review on the state-of-the art on bandwidth selection methods for regression discontinuity designs. Whenever a bandwidth choice is not manually supplied, the command \texttt{rdhte} internally relies on \texttt{rdbwhte} to implement data-driven bandwidth selection as a first step.

    \item \texttt{rdhte\_lincom}. This post-estimation command implements both estimation and robust bias corrected inference for linear combinations of the treatments effects estimated by \texttt{rdhte}, inspired by the built-in command {\tt lincom}, which cannot itself be used directly. The \texttt{Stata} post-estimation command \texttt{test} can be used directly for testing multiple linear hypotheses, but does not produce point estimates (e.g., for contrasts).
    
\end{itemize}

Section \ref{sec: Setup} introduces the sharp RD setup, highlighting the role of pretreatment covariates, both heterogeneity analysis (which is our main focus) and for efficiency improvements. We detail how heterogeneous treatment effects are recovered using local polynomial regression methods with semi-linear interactions. The methods are adapted from \cite{Calonico-Cattaneo-Farrell-Palomba-Titiunik_2025_wp}, and we refer to that paper for the relevant econometric theory. \cite{Cattaneo-Keele-Titiunik_2023_HandbookCh} offer a broader discussion on the role of covariate-adjustments in RD designs.

Within the heterogeneity framework of Section \ref{sec: Setup}, there are two distinct cases depending on the type of covariates used: (i) dummy variables (or, more generally, factor variables) that identify orthogonal (mutually exclusive) subgroups; or (ii) generic covariates (discrete or continuous). The first case, treated in detail in Section \ref{sec:subgroups}, arises when the covariates, and any included covariate interactions, correspond to non-overlapping subsets of the data, that is, when only binary orthogonal variables are used. This case brings the familiar subgroup analysis to RD designs, which naturally arises via indicator variables for different categories and from collapsing continuous variables into distinct bins. In this case, the treatment effect is different for each subgroup in an unrestricted way. 

In the second case (Section \ref{sec:general}), the covariates can be arbitrary, covering discrete, continuous, or mixed, and allowing for general interactions or transformations such as polynomials or other basis expansions thereof. For this generic setting, \cite{Calonico-Cattaneo-Farrell-Palomba-Titiunik_2025_wp} provide formal identification, estimation, and robust bias-corrected inference methods for heterogeneity analysis based on standard local polynomial regression methods with semi-linear interactions. The treatment effect heterogeneity is defined as a varying coefficient linear function of the pretreatment covariates. As an alternative, see \cite{Reguly2021-wp} and \cite{alcantara2025LearningConditionalAverage} for heterogeneous treatment effect estimation leveraging machine learning methods, and \cite{hsu2019testing,Hsu-Shen_2021_JAE} for specification testing methods. 

The package \texttt{rdhte} complements the popular package \texttt{rdrobust} \citep{Calonico-Cattaneo-Farrell-Titiunik_2017_Stata}, which focuses on estimation and robust bias-correction inference for the local average treatment effect at the cutoff in RD designs. Section \ref{sec: Comparison with rdrobust} compares the two RD packages, and explains precisely their differences. In certain specific cases, the package \texttt{rdhte} matches the analysis based on \texttt{rdrobust}.

Throughout Sections \ref{sec:subgroups}, \ref{sec:general}, and \ref{sec: Comparison with rdrobust}, we illustrate the features of {\tt rdhte} using the data from \cite{granzier2023coordination}, who studied coordination behavior in French two-round elections. We add to their results by studying treatment effect heterogeneity. The running variable is the vote margin in the first round election, and the outcome of interest is a binary indicator for running in the second round or not. We explore heterogeneity by party ideology and party strength, as defined in more detail below.

Section \ref{sec: Conclusion} concludes. Finally, the latest version of the package \texttt{rdhte}, replication files, and other related materials, are available at:
\begin{center}
    \href{https://rdpackages.github.io/rdhte/}{\texttt{https://rdpackages.github.io/rdhte/}}.
\end{center}

\section{Setup}\label{sec: Setup}

Extending RD analysis to heterogeneous treatment effects poses a nonparametric challenge, often requiring researchers to rely on semiparametric, yet parsimonious models. Following practice, \cite{Calonico-Cattaneo-Farrell-Palomba-Titiunik_2025_wp} examines the common approach of using local least squares regression with linear interactions, clarifying the conditions under which this method yields meaningful causal interpretations. The authors established that when potential outcomes follow a local linear-in-parameters, functional coefficient model, heterogeneous effects are identifiable and interpretable, particularly for binary orthogonal covariates (subgroup effects). \cite{Calonico-Cattaneo-Farrell-Palomba-Titiunik_2025_wp} further developed formal econometric methods for estimation and (robust bias-corrected) inference, including optimal bandwidth selection and standard error estimators robust to both heteroskedasticity and clustering. These results aim to add rigor and consistency to empirical practice, improving the applicability, and replicability, of RD heterogeneity analysis.

The observed random sample is $(Y_i, T_i, X_i, \bW_i', \bZ_i')'$, for $i=1,\dots,n$, where:
\begin{itemize}
    \item $Y_i = T_i Y_i(1) + (1-T_i) Y_i(0)$ is the observed outcome, and $Y_i(0)$ and $Y_i(1)$ are the underlying potential outcomes under control and treatment status, respectively;

    \item $T_i = \I(X_i \geq c)$ is the treatment assignment indicator with $X_i$ the continuous score and $c$ a known cutoff;

    \item $\bW_i$ is a $d$-dimensional vector of pretreatment covariates used for heterogeneity analysis; and

    \item $\bZ_i$ is a $d_z$-dimensional vector of pretreatment covariates used for efficiency improvements.
\end{itemize}

Without loss of generality, we set the cutoff $c=0$ to streamline the presentation. The canonical RD design corresponds to the case where neither $\bW_i$ nor $\bZ_i$ are included in the analysis.

\subsection{Average Treatment Effect \& Efficiency Covariates}

The canonical sharp RD average treatment effect (at the cutoff $X_i=c=0$) is
\begin{equation}
    \label{eq:ATE}
    \tau = \E[Y_i(1) - Y_i(0) | X_i = 0].
\end{equation}
It is common practice to employ the following least squares local polynomial RD estimator:
\begin{equation}
    \label{eq:ATE dot}
    \dot{\tau} = \be_{0}'\dot{\bbeta},
\end{equation}
where $\be_{\ell}$ denotes the conformable unit vector with a $1$ in its $(\ell+1)$th element, and
\begin{align}\label{eq:OLS fit -- RD}
    \left[\begin{array}{c}\dot{\balpha} \\ \dot{\bbeta} \end{array}\right]
    = \argmin_{\balpha,\bbeta}
      \sum_{i=1}^n \Big(Y_i - \mathbf{r}_p(X_i)'\balpha - T_i\mathbf{r}_p(X_i)'\bbeta
                   \Big)^2 K_h(X_i),
\end{align}
with $\mathbf{r}_p(u) = (1,u,\dots,u^p)'$ denoting the $p$th order polynomial expansion, and $K_h(u) = K(u/h)/h$ for a kernel (i.e., weighting) function $K(\cdot)$ and bandwidth $h$.

For the local-linear case ($p=1$), the underlying implementation employing the \texttt{Stata} command \texttt{reg} is
\begin{align}\label{eq: reg-cmd rdhte ATE}
    \text{\texttt{reg y t\#\#c.x}}
\end{align}
properly localized to $c=0$ and weighted using the kernel $K(\cdot)$. Then, $\dot{\tau}$ corresponds to the coefficient estimate associated with \texttt{t}.

The classical RD treatment effect estimator $\dot{\tau}$ is consistent for $\tau$ under standard regularity conditions. Similarly, $\be_{1}'\dot{\bbeta}$ is an estimator useful in the Kink RD design. \cite{Calonico-Cattaneo-Titiunik_2014_ECMA} studied MSE-optimal point estimation and robust bias-corrected inference for this case, while \cite{Calonico-Cattaneo-Farrell_2020_ECTJ} study optimal bandwidth selection for both point estimation and inference.

The covariates $\bZ_i$ can be included in the regression estimation to improve efficiency in the estimation of RD treatment effect $\tau$. \cite{Calonico-Cattaneo-Farrell-Titiunik_2019_RESTAT} studied MSE-optimal point estimation and robust bias-corrected inference for this case, and recommended the estimator
\begin{align*}
    \widetilde{\tau} = \be_{0}'\widetilde{\bbeta},
\end{align*}
where
\begin{align}\label{eq:OLS fit -- RD + cov.eff}
    \left[\begin{array}{c}\widetilde{\balpha} \\ \widetilde{\bbeta} \\ \widetilde{\bgamma} \end{array}\right]
    = \argmin_{\balpha,\bbeta,\bgamma}
      \sum_{i=1}^n \Big(Y_i - \mathbf{r}_p(X_i)'\balpha - T_i\mathbf{r}_p(X_i)'\bbeta - \bZ_i'\bgamma
                   \Big)^2 K_h(X_i).
\end{align}
The covariate-adjusted, possibly more efficient RD estimator $\widetilde{\tau}$ is consistent for $\tau$ under regularity conditions. The estimated coefficients $\widetilde{\bgamma}$ do not have a causal interpretation; they are fitted to improve the precision of $\widetilde{\bbeta}$. Standard Kink RD designs with covariate-adjustment for efficiency gains consider the estimator $\be_{1}'\widetilde{\bbeta}$. 

For the local-linear case ($p=1$), the underlying implementation employing the \texttt{Stata} command \texttt{reg} is
\begin{align}\label{eq: reg-cmd rdhte ATE + COV}
    \text{\texttt{reg y t\#\#c.x z}}
\end{align}
again properly localized to $c=0$ and weighted using the kernel $K(\cdot)$. Then, $\widetilde{\tau}$ corresponds to the coefficient associated with \texttt{t}.

The package \texttt{rdrobust} \citep{Calonico-Cattaneo-Farrell-Titiunik_2017_Stata} provides software implementation for estimation and inference on the RD average treatment effects, both with and without adding covariates for efficiency gains.

\subsection{Heterogeneous Average Treatment Effects}
    \label{sec:hte}

To complement prior literature focusing on the RD average treatment effect $\tau$, \cite{Calonico-Cattaneo-Farrell-Palomba-Titiunik_2025_wp} considers the (local to $X_i=c=0$) RD conditional average treatment effect (CATE) function
\begin{align*}
    \kappa(\bw) = \E[Y_i(1) - Y_i(0) | X_i = 0, \bW_i = \bw],
\end{align*}
which employs the covariates $\bW_i$ for RD treatment effect heterogeneity. The covariates $\bW_i$ are distinct from those used for efficiency improvements, $\bZ_i$, but importantly both are predetermined.

When $\bW_i$ is continuous and/or high-dimensional, the RD CATE function $\kappa(\bw)$ can be difficult to estimate nonparametrically without further restrictions. Thus, it is common practice to employ the semilinear least squares local polynomial estimation procedure:
\begin{equation}
    \label{eq:kappa hat}
    \widehat{\kappa}(\bw) = \be_{0}'\widehat{\bbeta} + \widehat{\bxi}'\begin{bmatrix}
           \mathbf{I}_d \\
           \mathbf{0}_{sd \times d}
         \end{bmatrix}\mathbf{w},
\end{equation}
where $\mathbf{I}_d$ denotes the $(d\times d)$ identity matrix, $\mathbf{0}_{sd \times d}$ denotes the $(sd\times d)$ matrix of zeros, $\bw$ takes values on the support of $\bW_i$, and
\begin{align}\label{eq:OLS fit -- RD + cov.hte}
    \left[\begin{array}{c}\widehat{\balpha} \\ \widehat{\blambda} \\ \widehat{\bxi} \end{array}\right]
    = \argmin_{\balpha,\bbeta,\blambda,\bxi}
        \sum_{i=1}^n
        \Big(Y_i &- \mathbf{r}_p(X_i)'\balpha
                   - T_i\mathbf{r}_p(X_i)'\bbeta \nonumber\\
                 &- [\mathbf{r}_s(X_i)\otimes \bW_i]'\blambda
                   - T_i[\mathbf{r}_s(X_i)\otimes \bW_i]'\bxi
        \Big)^2 K_h(X_i),
\end{align}
with $\otimes$ denoting the Kronecker product. \cite{Calonico-Cattaneo-Farrell-Palomba-Titiunik_2025_wp} study this estimation procedure and give easy-to-interpret sufficient conditions to ensure that $\widehat{\kappa}(\bw)$ is consistent for $\kappa(\bw)$. The most important of these is the assumption that the CATE function can be written as a (local) functional coefficient linear form: $\kappa(\bw) = \theta(x) + \bxi(x)'\bw$, see below. This is without loss of generality for binary orthogonal covariates (Section \ref{sec:subgroups}). \cite{Calonico-Cattaneo-Farrell-Palomba-Titiunik_2025_wp} also establish MSE-optimal bandwidth selection and point estimation, and valid robust bias-corrected inference for uncertainty quantification.

The generic form of $\widehat{\kappa}(\bw)$, and its underlying least squares fit \eqref{eq:OLS fit -- RD + cov.hte}, is notationally cumbersome but easy to understand. It arises from adding to the local polynomial estimation the interaction between the covariates $\bW_i$ and the polynomial approximation $\mathbf{r}_s(X_i)$ (which may be a different polynomial than the main effect). For example, if $p=s=1$, so that $\mathbf{r}_s(X_i)=(1,X_i)'$, and $\bW_i$ is a continuous variable, the {\tt Stata} implementation would be:
\begin{align}\label{eq: reg-cmd rdhte HTE}
    \text{\texttt{reg y t\#\#c.x\#\#c.w}}
\end{align}
properly localized to $c=0$ and weighted using the kernel $K(\cdot)$. The estimate $\widehat{\kappa}(w_0)$ of Equation \eqref{eq:kappa hat}, for a specific value $w_0$  is obtained as the coefficient on {\tt t} plus the product of $w_0$ and the coefficient on the interaction of {\tt t} and {\tt c.w}.

Covariate-adjustment based on $\bZ_i$ for efficiency gains is also allowed: following \cite{Calonico-Cattaneo-Farrell-Titiunik_2019_RESTAT}, in the regression \eqref{eq:OLS fit -- RD + cov.hte}, $\bZ_i$ and $\bZ_i\otimes\bW_i$ are included but without interaction with treatment assignment variable $T_i$. The resulting coefficients (on $\bZ_i$ and $\bZ_i\otimes\bW_i$) do not have a causal interpretation, but the resulting estimator $\widehat{\kappa}(\bw)$ can exhibit efficiency gains. We omit further details for brevity, but the local-linear command is:
\begin{align}\label{eq: reg-cmd rdhte HTE + COV}
    \text{\texttt{reg y t\#\#c.x\#\#c.w z\#\#c.w}}
\end{align}
properly localized to $c=0$ and weighted using the kernel $K(\cdot)$. The CATE estimate is obtained exactly as before.

The causal interpretation of the probability limit of the estimator $\widehat{\kappa}(\bw)$ must be carefully considered and depends on the type of covariates as follows.
\begin{itemize}
    \item Binary Orthogonal $\bW_i$. This case corresponds to situations where each component is binary, $\bW_i \in \{0,1\}^d$, and at most one component takes the value one, $\bW_i'\bW_i \leq 1$. By implication, the data is partitioned in disjoint groups as determined by $\bW_i$, and thus this case corresponds to subset analysis. The vector $\bW_i$ consists of indicator variables (for categories, levels, bins, etc) and their interactions. In this case, the estimator $\widehat{\kappa}(\bw)$ is fully nonparametric and consistent for generic $\kappa(\bw)$ under minimal regularity conditions, for all $\bw\in\{0,1\}^d$. In this special setting, the assumption $\kappa(\bw) = \theta(x) + \bxi(x)'\bw$ is without loss of generality, i.e., $\kappa(\bw)$ is unrestricted.

    \item Generic $\bW_i$. This case corresponds to situations where some components of $\bW_i$ are continuous, take several discrete values (which are \emph{not} ``dummied out''), or are mixed. In this case, the functional coefficient linear model is used by \cite{Calonico-Cattaneo-Farrell-Palomba-Titiunik_2025_wp} to restore a causal interpretation to the local least squares estimation procedure \eqref{eq:OLS fit -- RD + cov.hte}. Absent this assumption, $\widehat{\kappa}(\bw)$ is not consistent for a causal object, but rather its probability limit has the usual local best linear-in-parameters approximation interpretation. The identifying assumption is flexible in that it allows for nonlinear heterogeneity by including in $\bW_i$ transformations of the original covariates, such as polynomial expansions or other basis transformations.    
\end{itemize}

Regardless of the specific structure of $\bW_i$, \cite{Calonico-Cattaneo-Farrell-Palomba-Titiunik_2025_wp} develop MSE-optimal estimation and robust bias-correction inference for $\kappa(\bw)$, expanding prior results established for $\tau$. See \cite{Arai-Ichimura_2018_QE} and \cite{Calonico-Cattaneo-Farrell_2020_ECTJ} for more discussion on bandwidth selection for classical RD designs, \cite{Calonico-Cattaneo-Titiunik_2014_ECMA} and \citet{Calonico-Cattaneo-Farrell_2018_JASA,Calonico-Cattaneo-Farrell_2022_Bernoulli} for foundational theoretical analysis of robust bias-correction inference. \cite{Hyytinen-Tukiainen-etal2018_QE} and \cite{DeMagalhaes-etal_2025_PA} offer comprehensive empirical validation of those estimation and inference methods.

\section{Binary Orthogonal Covariates}
    \label{sec:subgroups}

We begin the illustration of heterogeneity analysis with {\tt rdhte} with the first case: binary orthogonal covariates (subgroup analysis). Recall the context of \cite{granzier2023coordination}, where the outcome is $Y_i \in \{0,1\}$ indicates running in the second round election, and the running variable $X_i$ is the vote margin in the first round. We will show the features of {\tt rdhte} using different measures of party ideology and strength. Unless noted otherwise, all results will use standard errors clustered by district (stored as {\tt cluster\_var}) following their original analysis. Because {\tt rdhte} is based on {\tt regress}, clustered standard errors are obtained using the standard {\tt vce} options: {\tt vce(cluster cluster\_var}) yields HC1 clustered standard errors, while {\tt vce(hc2 cluster\_var}) yields HC2, which is preferred. Currently, HC3 is not available, but it would provide a more robust option. We therefore recommend setting {\tt vce(hc2 cluster\_var}), which we use throughout. In other software, HC3 is available and thus set as the default for both clustered and independent data.

\subsection{Single Binary Variable}

The most basic form for such analysis is the case of a single binary variable, so that $d=1$ and $W_i \in \{0,1\}$. To illustrate, we create the dummy variable {\tt left}, taking the value one if the candidate belongs to the left or far left and zero otherwise. Running {\tt rdhte} yields a point estimate and robust bias corrected inference for each category.

{\fontsize{6}{7}\selectfont\begin{stlog}[auto]\input{output/out1a.log}\end{stlog}}

We see that the RD treatment effect is larger for left-of-center candidates  (0.089 versus 0.021) and is statistically significantly different from zero, which is not the case for the others. In this case, {\tt rdhte} automatically detects that {\tt w\_left} is an indicator for subgroups because {\tt w\_left} takes only the values zero and one. Any other coding (e.g., if it takes values 1 and 2) will require use of the {\tt i.}{\it varname} syntax. This output is omitted to save space but is available in the replication code.

The package, by default, automatically computes the optimal bandwidth for each group separately. The behavior can be overridden using the option {\tt bwjoint}. However, this is generally not advisable because the bias-variance trade-off that determines the optimal bandwidth need not be the same across subsets of the data, just as it need not be the same across different datasets. See the replication code for an example and the help file for {\tt rdbwhte} for more details on bandwidth selection.

A natural question following these results is whether the two subgroups are statistically significantly different. This is addressed using the featured post-estimation command {\tt rdhte\_lincom}, which computes the point estimate of the difference (in this case $\widehat{\kappa}(1) - \widehat{\kappa}(0)$) and the robust bias-corrected confidence interval and p-value. 

{\fontsize{8}{8}\selectfont\begin{stlog}[auto]\input{output/out1b.log}\end{stlog}}

We see that, indeed, the effect for left-of-center candidates is significantly higher at the 5\% level. In the next subsection, we also demonstrate post-estimation hypothesis testing using the Stata built-in command {\tt test}.

\subsection{``Dummied Out'' Factor Variable -- Unordered}

Moving beyond the case of a single binary variable, we now explore other instances where the set of covariates $\bW_i$ are binary orthogonal variables and therefore identify subgroups of the data. 

We obtain a more nuanced view of the heterogeneity by party ideology using the four-level factor variable {\tt w\_ideology}. In this case, the four-level factor is ``dummied out'' to obtain an indicator for each group. This happens automatically using the {\tt i.} syntax.

{\fontsize{6}{7}\selectfont\begin{stlog}[auto]\input{output/out2a.log}\end{stlog}}

For post-estimation, we use the built-in {\tt Stata} command {\tt test} to find that all the non-left categories are statistically indistinguishable from each other and from zero.

{\fontsize{8}{8}\selectfont\begin{stlog}[auto]\input{output/out2b.log}\end{stlog}}

\subsection{``Dummied Out'' Factor Variable -- Ordered}

We bring in another pre-treatment variable for heterogeneity analysis: the strength at the national level, defined as the average of first-round vote shares of all candidates of the same orientation at the national level. The raw measure is stored as {\tt w\_strength} and is considered in the next subsection. Here, we categorize strength into the four quartiles and obtain the effect for each. Compared to the case of {\tt w\_ideology}, this variable has a clear ordering. Note here that the {\tt i.} syntax is required; otherwise {\tt rdhte} will yield a linear fit treating the variable as continuous (see below).

{\fontsize{5}{7}\selectfont\begin{stlog}[auto]\input{output/out3.log}\end{stlog}}

We find that the causal effect is increasing with strength, an interesting if not surprising finding. This analysis depends on binning, and can be viewed intuitively (though not formally) as a four-piece approximation to an unknown, nonparametric $\kappa(w)$, following \cite{Cattaneo-Crump-Farrell-Feng_2024_AER}.

\subsection{Binary Interactions}

Finally, dummying out factor variables, and then interacting them, also yields binary orthogonal variables, one for each unique combination, provided the model is saturated. To illustrate, here we interact the dummy for left-of-center with an indicator for a ``strong'' candidate, defined as having above-median national strength.

{\fontsize{5}{7}\selectfont\begin{stlog}[auto]\input{output/out4.log}\end{stlog}}

As is typical for interaction effects, this adds nuance to the above findings, as we see that not only do left-of-center candidates have larger treatment effects, but this is even more pronounced when the candidates are stronger than the national median.

\section{Generic Covariates}
    \label{sec:general}

We now discuss the case of generic covariates: $\bW_i$ is not a set of binary orthogonal variables, and so instead of obtaining heterogeneity by subgroups, we obtain a linear-in-parameters estimate of the function $\kappa(\bw)$. As discussed above, \cite{Calonico-Cattaneo-Farrell-Palomba-Titiunik_2025_wp} use a functional coefficient model assumption to obtain a causal interpretation of the probability limit of $\widehat{\kappa}(\bw)$. In this case, instead of selecting an optimal bandwidth for each subset, a single bandwidth is selected.

\subsection{Continuous Variables and Replication with Linear Regression}

To illustrate we use the measure of strength, {\tt w\_strength}. Before, when grouped by quantiles, we saw that the treatment effect was increasing in strength. It may be that a linear fit is an appropriate and parsimonious way of capturing this relationship. Using {\tt w\_strength}, we obtain the following. For later comparison with results from {\tt regress}, here we use the uniform kernel to obtain an unweighted local least squares fit. Note that the output of {\tt rdhte} is different relative to the subset analysis, reflecting that there is only one bandwidth and the coefficients here represent qualitatively different objects. 

{\fontsize{8}{8}\selectfont\begin{stlog}[auto]\input{output/out5a.log}\end{stlog}}

In Section \ref{sec:subgroups}, each row of the output table showed the treatment effect estimate (and robust bias-corrected inference) for each subset of the data. Here, instead, we have the ``intercept'' and ``slope'' terms of the estimate $\widehat{\kappa}(\bw)$ of Equation \eqref{eq:kappa hat}. That is, for a specific level of candidate strength $w_0$, we would obtain $\widehat{\kappa}(w_0) = -0.055 + 0.262 \times w_0$. The value -0.055 represents $\widehat{\kappa}(0)$, which may or may not have conceptual meaning, depending on the context and the definition of $\bW_i$. 

To build intuition, this interpretation is identical to interpreting coefficients in a linear regression, because here (by default) we are using {\tt rdhte} with a polynomial order of $p=s=1$. Recall that identical estimates can be obtained using {\tt regress}, properly localized and weighted, as shown in Equation \eqref{eq: reg-cmd rdhte HTE}. Here, we have used the uniform kernel, so we must only localize using the same bandwidth in order to obtain the same point estimates. We extract the bandwidth from the ereturns of {\tt rdhte}, generate the treatment indicator, and then run the regression.

{\fontsize{8}{8}\selectfont\begin{stlog}[auto]\input{output/out5b.log}\end{stlog}}

Here, the relevant coefficients are those on {\tt 1.T} and {\tt T\#c.w\_strength}. 

It is important to remember that this only replicates the point estimates, for \emph{inference}, robust bias correction is required. The inference measures (standard errors, t-statistics, confidence intervals, and p-values) in this {\tt regress} output are not valid.

\subsection{Interactions}

As a final illustration, consider the analog of the binary interaction at the close of Section \ref{sec:subgroups}. There, we obtained an estimate for each of the four categories reflecting left or not and above/below median strength. Here, we consider the interaction of the binary indicator {\tt w\_left} with the continuous measure of strength. By fully saturating the model, we obtain a separate intercept and slope for left-of-center and for center-and-right candidates. 

{\fontsize{7}{8}\selectfont\begin{stlog}[auto]\input{output/out6.log}\end{stlog}}

This output reveals the same qualitative conclusion as the binary interaction, but expressed in a different way. It is worth noting that, because this model is fully saturated, the same results can be obtained studying {\tt w\_strength} separately for each category of {\tt w\_left}. This is shown in the replication code, but omitted here to save space. 

\section{Comparison with {\tt rdrobust}}\label{sec: Comparison with rdrobust}

In this section, we briefly illustrate how {\tt rdhte} does, and does not, match the output from the popular command {\tt rdrobust} \citep{Calonico-Cattaneo-Farrell-Titiunik_2017_Stata}. For simplicity, all analyses in this section do not use clustered standard errors, and so for {\tt rdhte} the HC3 option is used. We show replication of average and of heterogeneous treatment effects. Both packages can use covariates for efficiency (using the {\tt covs\_eff} option in {\tt rhdte}), but this is not a point of comparison here.

\subsection{Implementation Differences}

The discrepancies arise for three main reasons, all of which are driven by the fact that {\tt rdhte} (in all platforms) relies on built-in base commands for least squares regression.
\begin{enumerate}
    \item {\tt rdrobust} allows for, and by default uses, a different bandwidth for bias correction, $b$, than for the main estimation, $h$, with $\rho = h / b$. In contrast, {\tt rdhte} forces $h=b$, i.e. $\rho=1$. This choice has known optimality properties and is reliable in practice \citep{Calonico-Cattaneo-Farrell_2020_ECTJ,Calonico-Cattaneo-Farrell_2022_Bernoulli}.

    \item {\tt rdrobust} uses a custom nearest-neighbor variance estimator by default, and also allows for HC0--HC3, as well as for HC and NN cluster-robust variance estimators. In contrast, {\tt rdhte} uses HC3 by default, and also allows for other variance estimators (heteroskedasticity-robust and cluster-robust) available via the corresponding base least squares command (when applicable).

    \item {\tt rdrobust} takes a two-sample approach for estimation and inference, while {\tt rdhte} takes a one-sample approach (i.e., single linear regression fit with interactions). This discrepancy in implementation approaches leads to different degrees of freedom adjustments, which in some cases can lead to slightly different standard error estimators (e.g., when implementing HC1). In addition, this discrepancy can lead to different matrix inversions and, by implication, potentially different covariates may be dropped due to multi-collinearity.
\end{enumerate}

\subsection{Average Treatment Effect}

First, consider the RD average treatment effect $\tau$ of Equation \eqref{eq:ATE}. 

Both {\tt rdrobust} and {\tt rdhte} can be used to obtain the corresponding estimator $\dot{\tau}$ of \eqref{eq:ATE dot} along with robust bias-corrected inference. For {\tt rdhte}, the average is estimated if no heterogeneity variables are specified. However, the default implementations differ, and so the output will not match, as shown here.

{\fontsize{8}{8}\selectfont\begin{stlog}[auto]\input{output/out7a.log}\end{stlog}}

If all these settings are synchronized, the two commands report the same results (up to numerical differences in computation). We illustrate this point by setting $h=0.1$ (which automatically forces $h=b$ or $\rho=1$ in {\tt rdrobust}), and variance estimation to HC3.

{\fontsize{8}{8}\selectfont\begin{stlog}[auto]\input{output/out7b.log}\end{stlog}}

\subsection{Subgroup Analysis}

The same equivalence can be obtained when conducting heterogeneity analysis by subgroup, because this amounts to estimating the average effect for a subset of the data. Here we illustrate that, again enforcing common settings, {\tt rdrobust} obtains the same results as {\tt rdhte} for left-of-center candidates (the other group is omitted to save space).

{\fontsize{6}{7}\selectfont\begin{stlog}[auto]\input{output/out7c.log}\end{stlog}}

\section{Conclusion}\label{sec: Conclusion}

This article illustrated the main functionalities of the package {\tt rdhte} for heterogeneous RD treatment effects estimation and robust bias-corrected inference. We also discussed how this package complements the popular RD package {\tt rdrobust}. The methods implemented by the package {\tt rdhte} can also be used in the context of multi-cutoff and multi-score RD designs by discretizing the analysis along the multi-dimensional assignment rule; see \citet{Cattaneo-Titiunik-VazquezBare_2020_Stata} for more discussion. Furthermore, {\tt rdhte} can also be used to implement other RD designs involving comparisons across subgroups, such as in difference-in-cutoffs designs \citep{Grembi-Nannicini-Troiano_2016_AEJ-Applied} or dynamic designs \citep{Hsu-Shen_2024_QE}. We do not discuss these connections further to conserve space.

\section{Acknowledgments}

Cattaneo and Titiunik gratefully acknowledge financial support from the National Science Foundation through grants SES-2019432 and SES-2241575.

\bibliographystyle{sj}
\bibliography{CCFPT_2025_Stata--bib}

\ifnum 23=1 \def\bibname{Reference}
\else \def\bibname{References} \fi
\begin{thebibliography}{23}
\expandafter\ifx\csname natexlab\endcsname\relax\def\natexlab#1{#1}\fi
\expandafter\ifx\csname url\endcsname\relax
  \def\url#1{\texttt{#1}}\fi
\expandafter\ifx\csname urlprefix\endcsname\relax\def\urlprefix{URL }\fi

\bibitem[{Alcantara et~al.(2025)Alcantara, Hahn, Carvalho, and
  Lopes}]{alcantara2025LearningConditionalAverage}
Alcantara, R., P.~R. Hahn, C.~Carvalho, and H.~Lopes. 2025.
\newblock Learning {{Conditional Average Treatment Effects}} in {{Regression
  Discontinuity Designs}} Using {{Bayesian Additive Regression Trees}}.
\newblock \emph{arXiv preprint arXiv:2503.00326} .

\bibitem[{Arai and Ichimura(2018)}]{Arai-Ichimura_2018_QE}
Arai, Y., and H.~Ichimura. 2018.
\newblock Simultaneous Selection of Optimal Bandwidths for the Sharp Regression
  Discontinuity Estimator.
\newblock \emph{Quantitative Economics} 9(1): 441--482.

\bibitem[{Calonico et~al.(2018)Calonico, Cattaneo, and
  Farrell}]{Calonico-Cattaneo-Farrell_2018_JASA}
Calonico, S., M.~D. Cattaneo, and M.~H. Farrell. 2018.
\newblock On the Effect of Bias Estimation on Coverage Accuracy in
  Nonparametric Inference.
\newblock \emph{Journal of the American Statistical Association} 113(522):
  767--779.

\bibitem[{Calonico et~al.(2020)Calonico, Cattaneo, and
  Farrell}]{Calonico-Cattaneo-Farrell_2020_ECTJ}
\mbox{\vrule width30.25006ptheight2.62222ptdepth-2.25222pt}. 2020.
\newblock Optimal Bandwidth Choice for Robust Bias Corrected Inference in
  Regression Discontinuity Designs.
\newblock \emph{Econometrics Journal} 23(2): 192--210.

\bibitem[{Calonico et~al.(2022)Calonico, Cattaneo, and
  Farrell}]{Calonico-Cattaneo-Farrell_2022_Bernoulli}
\mbox{\vrule width30.25006ptheight2.62222ptdepth-2.25222pt}. 2022.
\newblock Coverage Error Optimal Confidence Intervals for Local Polynomial
  Regression.
\newblock \emph{Bernoulli} 28(4): 2998--3022.

\bibitem[{Calonico et~al.(2025)Calonico, Cattaneo, Farrell, Palomba, and
  Titiunik}]{Calonico-Cattaneo-Farrell-Palomba-Titiunik_2025_wp}
Calonico, S., M.~D. Cattaneo, M.~H. Farrell, F.~Palomba, and R.~Titiunik. 2025.
\newblock Treatment Effect Heterogeneity in Regression Discontinuity Designs.
\newblock \emph{arXiv preprint arXiv:2503.13696} .

\bibitem[{Calonico et~al.(2017)Calonico, Cattaneo, Farrell, and
  Titiunik}]{Calonico-Cattaneo-Farrell-Titiunik_2017_Stata}
Calonico, S., M.~D. Cattaneo, M.~H. Farrell, and R.~Titiunik. 2017.
\newblock \texttt{rdrobust}: Software for Regression Discontinuity Designs.
\newblock \emph{Stata Journal} 17(2): 372--404.

\bibitem[{Calonico et~al.(2019)Calonico, Cattaneo, Farrell, and
  Titiunik}]{Calonico-Cattaneo-Farrell-Titiunik_2019_RESTAT}
\mbox{\vrule width30.25006ptheight2.62222ptdepth-2.25222pt}. 2019.
\newblock Regression Discontinuity Designs using Covariates.
\newblock \emph{Review of Economics and Statistics} 101(3): 442--451.

\bibitem[{Calonico et~al.(2014)Calonico, Cattaneo, and
  Titiunik}]{Calonico-Cattaneo-Titiunik_2014_ECMA}
Calonico, S., M.~D. Cattaneo, and R.~Titiunik. 2014.
\newblock Robust Nonparametric Confidence Intervals for
  Regression-Discontinuity Designs.
\newblock \emph{Econometrica} 82(6): 2295--2326.

\bibitem[{Cattaneo et~al.(2024)Cattaneo, Crump, Farrell, and
  Feng}]{Cattaneo-Crump-Farrell-Feng_2024_AER}
Cattaneo, M.~D., R.~K. Crump, M.~H. Farrell, and Y.~Feng. 2024.
\newblock On Binscatter.
\newblock \emph{American Economic Review} 114(5): 1488--1514.

\bibitem[{Cattaneo et~al.(2019)Cattaneo, Idrobo, and
  Titiunik}]{Cattaneo-Idrobo-Titiunik2019_book}
Cattaneo, M.~D., N.~Idrobo, and R.~Titiunik. 2019.
\newblock \emph{A Practical Introduction to Regression Discontinuity Designs:
  Foundations}.
\newblock Cambridge Elements: Quantitative and Computational Methods for Social
  Science, Cambridge University Press.

\bibitem[{Cattaneo et~al.(2023{\natexlab{a}})Cattaneo, Idrobo, and
  Titiunik}]{Cattaneo-Idrobo-Titiunik2023_book}
\mbox{\vrule width30.25006ptheight2.62222ptdepth-2.25222pt}.
  2023{\natexlab{a}}.
\newblock \emph{A Practical Introduction to Regression Discontinuity Designs:
  Extensions}.
\newblock Cambridge Elements: Quantitative and Computational Methods for Social
  Science, Cambridge University Press, arXiv:2301.08958.

\bibitem[{Cattaneo et~al.(2023{\natexlab{b}})Cattaneo, Keele, and
  Titiunik}]{Cattaneo-Keele-Titiunik_2023_HandbookCh}
Cattaneo, M.~D., L.~Keele, and R.~Titiunik. 2023{\natexlab{b}}.
\newblock Covariate Adjustment in Regression Discontinuity Designs.
\newblock In \emph{Handbook of Matching and Weighting in Causal Inference}, ed.
  D.~S.~S. J.~R.~Zubizarreta, E. A.~Stuart and P.~R. Rosenbaum, chap.~8,
  153--168. Boca Raton, FL: Chapman \& Hall.

\bibitem[{Cattaneo and Titiunik(2022)}]{Cattaneo-Titiunik_2022_ARE}
Cattaneo, M.~D., and R.~Titiunik. 2022.
\newblock Regression Discontinuity Designs.
\newblock \emph{Annual Review of Economics} 14: 821--851.

\bibitem[{Cattaneo et~al.(2020)Cattaneo, Titiunik, and
  Vazquez-Bare}]{Cattaneo-Titiunik-VazquezBare_2020_Stata}
Cattaneo, M.~D., R.~Titiunik, and G.~Vazquez-Bare. 2020.
\newblock Analysis of Regression Discontinuity Designs with Multiple Cutoffs or
  Multiple Scores.
\newblock \emph{Stata Journal} 20(4): 866--891.

\bibitem[{De~Magalh{\~a}es et~al.(2025)De~Magalh{\~a}es, Hangartner, Hirvonen,
  Meril{\"a}inen, Ruiz, and Tukiainen}]{DeMagalhaes-etal_2025_PA}
De~Magalh{\~a}es, L., D.~Hangartner, S.~Hirvonen, J.~Meril{\"a}inen, N.~A.
  Ruiz, and J.~Tukiainen. 2025.
\newblock When Can We Trust Regression Discontinuity Design Estimates from
  Close Elections? Evidence from Experimental Benchmarks.
\newblock \emph{Political Analysis} .

\bibitem[{Granzier et~al.(2023)Granzier, Pons, and
  Tricaud}]{granzier2023coordination}
Granzier, R., V.~Pons, and C.~Tricaud. 2023.
\newblock Coordination and Bandwagon Effects: How Past Rankings Shape the
  Behavior of Voters and Candidates.
\newblock \emph{American Economic Journal: Applied Economics} 15(4): 177--217.

\bibitem[{Grembi et~al.(2016)Grembi, Nannicini, and
  Troiano}]{Grembi-Nannicini-Troiano_2016_AEJ-Applied}
Grembi, V., T.~Nannicini, and U.~Troiano. 2016.
\newblock Do fiscal rules matter?
\newblock \emph{American Economic Journal: Applied Economics} 8(3): 1--30.

\bibitem[{Hsu and Shen(2019)}]{hsu2019testing}
Hsu, Y.-C., and S.~Shen. 2019.
\newblock Testing treatment effect heterogeneity in regression discontinuity
  designs.
\newblock \emph{Journal of Econometrics} 208(2): 468--486.

\bibitem[{Hsu and Shen(2021)}]{Hsu-Shen_2021_JAE}
\mbox{\vrule width30.25006ptheight2.62222ptdepth-2.25222pt}. 2021.
\newblock Testing Monotonicity of Conditional Treatment Effects under
  Regression Discontinuity Designs.
\newblock \emph{Journal of Applied Econometrics} 36(3): 346--366.

\bibitem[{Hsu and Shen(2024)}]{Hsu-Shen_2024_QE}
\mbox{\vrule width30.25006ptheight2.62222ptdepth-2.25222pt}. 2024.
\newblock Dynamic regression discontinuity under treatment effect
  heterogeneity.
\newblock \emph{Quantitative Economics} 15(4): 1035--1064.

\bibitem[{Hyytinen et~al.(2018)Hyytinen, Meril{\"a}inen, Saarimaa, Toivanen,
  and Tukiainen}]{Hyytinen-Tukiainen-etal2018_QE}
Hyytinen, A., J.~Meril{\"a}inen, T.~Saarimaa, O.~Toivanen, and J.~Tukiainen.
  2018.
\newblock When does regression discontinuity design work? Evidence from random
  election outcomes.
\newblock \emph{Quantitative Economics} 9(2): 1019--1051.

\bibitem[{Reguly(2021)}]{Reguly2021-wp}
Reguly, {\'A}. 2021.
\newblock Heterogeneous Treatment Effects in Regression Discontinuity Designs.
\newblock \emph{arXiv preprint arXiv:2106.11640} .

\end{thebibliography}

\section{About the Authors}

\noindent
Sebastian Calonico is an Assistant Professor at the University of California at Davis.

\noindent
Matias D. Cattaneo is a Professor at Princeton University.

\noindent
Filippo Palomba is a Ph.D. candidate at Princeton University.

\noindent
Max H. Farrell is an Associate Professor at the University of California at Santa Barbara.

\noindent
Rocio Titiunik is a Professor at Princeton University.

\end{document}